\documentclass{aa}
\usepackage{epsf}
\begin{document}

%
   \title{Density waves in the shearing sheet }
   \subtitle{IV. Interaction with a live dark halo}

   \author{B. Fuchs}


   \institute{Astronomisches Rechen--Institut,
              M\"onchhofstrasse 12--14, 69120 Heidelberg, Germany}

   \date{Received 2003; accepted}

   \abstract{
It is shown that if the self--gravitating shearing sheet, a model of a patch of 
a galactic disk, is embedded in a live
dark halo, this has a strong effect on the dynamics of density waves in the
sheet. I describe how the density waves and the halo interact via halo particles
either on orbits in resonance with the wave or on non-resonant orbits. Contrary
to expectation the presence of the halo leads to a very considerable
enhancement of the amplitudes of the density waves in the shearing sheet. This
effect appears to be the equivalent of the recently reported enhanced growth of 
bars in numerically simulated stellar disks embedded in live dark halos.
Finally I discuss the transfer of linear momentum from a density wave in the 
sheet to the halo and show that it is mediated only by halo particles on 
resonant orbits.
      }
		
   \mail{fuchs@ari.uni-heidelberg.de}
    		
   \maketitle

%

\section{Introduction}

The shearing sheet (Goldreich \& Lynden--Bell 1965, Julian \& Toomre 
1966) model has been developed as a tool to study the dynamics of
galactic disks and is particularly well suited to describe theoretically
the dynamical mechanisms responsible for the formation of spiral arms.
For the sake of simplicity the model describes only the dynamics of a 
patch of a galactic disk. It is assumed to be infinitesimally thin and 
its radial size is assumed to be much smaller than the disk. Polar 
coordinates can be therefore rectified to pseudo-Cartesian
coordinates and the velocity field of the differential rotation of the
disk can be approximated by a linear shear flow. These simplifications
allow an analytical treatment of the problem, which helps to clarify 
the underlying physical processes operating in the disk. 

In the present paper of this series I discuss the dynamical effects if the
shearing sheet is immersed in a live dark halo. Dark halos are usually thought
to stabilize galactic disks against non-axisymmetric instabilities. This was
first proposed by Ostriker \& Peebles (1973) on the basis of
-- low--resolution -- numerical simulations. Their physical argument was that
the presence of a dark halo reduces the destabilizing self--gravity of the
disks. Doubts about an entirely passive role of dark halos were raised by 
Toomre (1977), but he (Toomre 1981) also pointed out that a dense core of a 
dark halo may cut the feed--back loop of the corotation amplifier of bars or
spiral density waves and suppress thus their growth. Recent high-resolution 
numerical simulations by Debattista \& Sellwood (2000) and Athanassoula 
(2002, 2003) have shown that quite the reverse, a {\em destabilization} of
disks immersed in dark halos, might be actually true. Athanassoula (2002) 
demonstrated clearly that much stronger bars grow in the simulations 
if the disk is embedded in a live dark halo instead of a static halo potential.
This is attributed to angular momentum transfer from the bar to the halo via 
halo particles on resonant orbits. Angular momentum exchange between disk and 
halo has been addressed since the pioneering work of Weinberg (1985) in many 
studies theoretically or by numerical simulations and I refer to Athanassoula 
(2003) for an overview of the literature. Toomre (1981) has shown how the bar
instability can be understood as an interference of spiral density waves in a
resonance cavity between the corotation amplifier and an inner reflector
(cf.~also Fuchs 2004). Thus it is to be expected that a live dark halo will be 
also responsive to spiral density waves and develop wakes, which I investigate
here using the shearing sheet model.

\section{Boltzmann equations}

The evolution of the distribution function of the disk stars in phase space is
described by the linearized 4--dimensional Boltzmann equation
\begin{eqnarray}
&&
\frac{\partial f_{\rm d1}}{\partial t} + u \frac{\partial f_{\rm d1}}
{\partial x} + v \frac{\partial f_{\rm d1}}{\partial y} \nonumber \\ &&
- \frac{\partial \Phi_{\rm d0}+\Phi_{\rm h0}}{\partial x}\frac{\partial
f_{\rm d1}}{\partial u} 
- \frac{\partial \Phi_{\rm d0}+\Phi_{\rm h0}}{\partial y}
\frac{\partial f_{\rm d1}}{\partial v} \nonumber \\ &&
 -\frac{\partial \Phi_{\rm d1}+\Phi_{\rm h1}}{\partial x}\frac{\partial
 f_{\rm d0}}{\partial u} 
  - \frac{\partial \Phi_{\rm d1}+\Phi_{\rm h1}}{\partial y}
\frac{\partial f_{\rm d0}}{\partial v} =0 \,,
\end{eqnarray}
where ($x$, $y$) denote spatial coordinates with $y$ pointing in the direction 
of galactic rotation and ($u$, $v$) are the corresponding velocity components,
respectively. Equation (1) has been derived from the general 6--dimensional
Boltzmann equation assuming delta--function like dependencies of the
distribution function on the vertical $z$ coordinate and the vertical $w$
velocity component, respectively, and integrating the Boltzmann equation with
respect to them. A perturbation Ansatz of the form
\begin{equation}
f_{\rm d} = f_{\rm{d0}} + f_{\rm{d1}} \,, \; 
\Phi_{\rm d} = \Phi_{\rm{d0}} + \Phi_{\rm{d1}}\,, \; 
\Phi_{\rm h} = \Phi_{\rm{h0}} + \Phi_{\rm{h1}} 
\end{equation}
is chosen for the distribution function of the disk stars and the
gravitational potentials of the disk and the halo, respectively, and the 
Boltzmann equation (1) has been linearized accordingly. 

Similarly the linearized Boltzmann equation for the halo particles can be 
written as
\begin{eqnarray}
&&
\frac{\partial f_{\rm h1}}{\partial t} + u \frac{\partial f_{\rm h1}}
{\partial x} + v \frac{\partial f_{\rm h1}}{\partial y}
+ w \frac{\partial f_{\rm h1}}{\partial z} \nonumber \\ &&
- \frac{\partial \Phi_{\rm d0}+\Phi_{\rm h0}}{\partial x}\frac{\partial
f_{\rm h1}}{\partial u}- 
\frac{\partial \Phi_{\rm d0}+\Phi_{\rm h0}}{\partial y}\frac{\partial
f_{\rm h1}}{\partial v} \nonumber \\ && -\frac{\partial \Phi_{\rm d0}+
\Phi_{\rm h0}}{\partial z}\frac{\partial f_{\rm h1}}{\partial w} 
-\frac{\partial \Phi_{\rm h1}+\Phi_{\rm d1}}{\partial x}
\frac{\partial f_{\rm h0}}{\partial u} \nonumber \\ &&
 - \frac{\partial \Phi_{\rm h1}+\Phi_{\rm d1}}{\partial y}
\frac{\partial f_{\rm h0}}{\partial v} 
 - \frac{\partial \Phi_{\rm h1}+\Phi_{\rm d1}}{\partial z}
\frac{\partial f_{\rm h0}}{\partial w} = 0 \,.
\end{eqnarray}
The choice of the dark halo model is lead by the following considerations. One
of the deeper reasons for the success of the infinite shearing sheet model to 
describe spiral density waves realistically is the rapid convergence of the 
Poisson integral in self--gravitating disks (Julian \& Toomre 1966). Consider, 
for example, the potential of a sinusoidal density perturbation
\begin{equation} 
\Phi(x,y) = - G \int_{-\infty}^{\infty} dx' \int_{-\infty}^{\infty} dy'
\frac{\Sigma_{10} \sin{(kx')}}{\sqrt{ (x-x')^2 + (y-y')^2}} \,,
\end{equation} 
where $G$ denotes the constant of gravitation. At $x=0$
\begin{eqnarray} 
&& \Phi = - 4G \Sigma_{10} \sin{(kx)} \lim_{ x_{\rm L} \to 
\infty} \frac{{\rm Si}(k x_{\rm L})}{k} = \nonumber \\ &&
 - \frac{2 \pi G \Sigma_{10}\sin{(kx)}}{k}\,.
\end{eqnarray}
The sine integral in equation (5) converges so rapidly that it reaches at
$k x_{\rm L} = \frac{\pi}{2}$ already 87\% of its asymptotic value. Thus
the ``effective range'' of gravity is about only a quarter of a wave length. The
shearing sheet models effectively patches of galactic disks of such size. The
wave lengths of density waves are of the order of the critical wave length 
(Julian \& Toomre 1966)
\begin{equation} 
\lambda_{\rm crit} = \frac{4 \pi^2 G \Sigma_{\rm d}}{\kappa^2} \,,
\end{equation} 
where $\kappa$ denotes the epicyclic frequency of the stellar orbits and
$\Sigma_{\rm d}$ is the surface density of the sheet. In the solar neighbourhood
in the Milky Way, for instance, the critical wave length is $\lambda_{\rm
crit}=5$ kpc. Thus it is reasonable to neglect over such length scales, like in
the shearing sheet model, the curvature of the mean circular orbits of the 
stars around the galactic center or the gradient of the surface density. The 
curvature of the stellar orbits due to the epicyclic motions of the stars, 
on the other hand, cannot be neglected and is indeed not neglected in the
shearing sheet model. The radial size of an epicycle is approximately given by
$\sigma_{\rm u}/\kappa$, where $\sigma_{\rm u}$ denotes the radial velocity
dispersion of the stars, and the ratio of epicycle size and critical wave 
length is given by
\begin{equation} 
\frac{\sigma_{\rm u}}{\kappa \lambda_{\rm crit}} = 0.085 Q
\end{equation} 
in terms of the Toomre stability parameter $Q$ (Toomre 1964) which is typically
of the order of 1 to 2. Concurrent to these approximations I assume a dark halo 
which is homogeneous in its unperturbed state. Accordingly the curvature of the
unperturbed orbits of the halo particles is neglected on the scales considered
here and the particles are assumed to be on straight--line orbits. The equations
of motion of the halo particles are the characteristics of the Boltzmann 
equation. In a homogeneous halo $\ddot{\bf x}=\nabla (\Phi_{\rm d0}+
\Phi_{\rm h0}) = 0$, and in accordance with this assumption I neglect the force
terms $\nabla \Phi_{\rm d0}$ and $\nabla \Phi_{\rm h0}$ in the Boltzmann 
equation (3). This simplifies its solution considerably. The disadvantage of 
such a model is that there are no higher--order resonances of the 
orbits of the halo particles with the density waves as described by Weinberg 
(1985) or observed in the high--resolution simulations by Athanassoula (2002, 
2003). However their effect was shown to be much less important than the main 
resonances of the particles with the density waves, which are properly 
described in the present model.  

\section{Halo dynamics}

The Boltzmann equation (3) can be viewed as a linear partial differential 
equation for the perturbation of the distribution function of the halo 
particles, $f_{\rm  h1}$, with inhomogeneities depending on the perturbations 
of the gravitational potentials of the disk and the halo, $\nabla 
\Phi_{\rm d1}$ and $\nabla \Phi_{\rm h1}$, respectively. Thus the equation 
can be solved for the disk and halo inhomogeneities separately and the 
solutions combined afterwards by superposition.

\subsection{Jeans instability of the halo}

I consider first that part of the Boltzmann equation (3)
\begin{eqnarray}
&&
\frac{\partial f_{\rm h1}}{\partial t} + u \frac{\partial f_{\rm h1}}
{\partial x} + v \frac{\partial f_{\rm h1}}{\partial y}
+ w \frac{\partial f_{\rm h1}}{\partial z} \\ &&
-\frac{\partial \Phi_{\rm h1}}{\partial x}\frac{\partial f_{\rm h0}}
{\partial u} \nonumber  - \frac{\partial \Phi_{\rm h1}}{\partial y}
\frac{\partial f_{\rm h0}}{\partial v} 
 - \frac{\partial \Phi_{\rm h1}}{\partial z}
\frac{\partial f_{\rm h0}}{\partial w} = 0 \nonumber \,.
\end{eqnarray}
Equation (8) is Fourier transformed with respect to time and spatial
coordinates and I consider in the following Fourier terms
\begin{equation}
f_{\rm {h\bf k},\omega},\,\Phi_{\rm {h\bf k},\omega} \times
e^{i(\omega t + ({\bf k},{\bf x}))}
\end{equation}
with frequency $\omega$ and wave vector ${\bf k}$. If orthogonal spatial
coordinates $\xi$, $\eta,$ and $\zeta$ with $\xi$ parallel to the wave vector
$\bf{k}$ are introduced, equation (8) takes the form
\begin{equation}
i \omega f_{\rm h{\bf k}} + \upsilon i k f_{\rm h{\bf k}} - i k 
\Phi_{\rm h{\bf k}} \frac{\partial f_{\rm h0}}{\partial \upsilon} = 0 \,,
\end{equation}
where $k = |{\bf k}|$ and $\upsilon$ denotes the velocity component
parallel to the $\xi$-axis. Equation (10) can be immediately integrated with
respect to the velocity components perpendicular to ${\bf k}$. Adopting for the 
background distribution function a Gaussian distribution, $ f_{\rm h0} = 
\frac{\rho_{\rm b}}{\sqrt{2 \pi}\sigma_{\rm h}}\exp{-\frac{\upsilon^2}{2
\sigma_{\rm h}^2}}$, the solution of equation (10) is given by
\begin{equation}
f_{\rm h{\bf k}} = - \Phi_{\rm h{\bf k}} \frac{\upsilon}{\sigma_{\rm h}^2}
\frac{k}{\omega + k \upsilon}\frac{\rho_{\rm b}}
{\sqrt{2 \pi}\sigma_{\rm h}}\exp{-\frac{\upsilon^2}{2 \sigma_{\rm h}^2}}\,.
\end{equation}
The corresponding Fourier coefficient of the density perturbation of the halo
particles is
\begin{equation}
\rho_{\rm h{\bf k}} = - \Phi_{\rm h{\bf k}}\frac{\rho_{\rm b}}
{\sqrt{2 \pi}\sigma_{\rm h}^3} \int_{-\infty}^\infty d \upsilon
\frac{\upsilon}{\frac{\omega}{k} + \upsilon}\exp{-\frac{\upsilon^2}{2 
\sigma_{\rm h}^2}}\,.
\end{equation}
The perturbations of the halo are supposed to be self--gravitating and the
gravitational potential has to solve the Poisson equation
\begin{eqnarray}
&&  \{ \frac{\partial^2}{\partial x^2} +
\frac{\partial^2}{\partial y^2} + \frac{\partial^2}{\partial z^2} \}
\Phi_{\rm h{\bf k}} e^{i({\bf k},{\bf x})} =\\ &&
- k^2 \Phi_{\rm h{\bf k}} e^{i({\bf k},{\bf x})} = 4 \pi G \rho_{\rm h{\bf k}} 
 e^{i({\bf k},{\bf x})} \,. \nonumber
\end{eqnarray}
Combining equations (12) and (13) leads to the dispersion relation
\begin{equation}
k^2 = \frac{4 \pi G \rho_{\rm b}}
{\sqrt{2 \pi}\sigma_{\rm h}^3} \int_{-\infty}^\infty d \upsilon
\frac{\upsilon}{\frac{\omega}{k} + \upsilon}\exp{-\frac{\upsilon^2}{2 
\sigma_{\rm h}^2}}\,,
\end{equation}
which is well known from plasma physics (cf. Kegel 1998). Indeed, the
dispersion relation (14) describes simply the Jeans collapse of the halo. From
the imaginary part of equation (14) one concludes $\Re(\omega) = 0$ and using
formula 3.466 of Gradshteyn \& Ryzhik (2000) one obtains from its real part the
dispersion relation
\begin{equation}
k^2 = \frac{4 \pi G \rho_{\rm b}}
{\sigma_{\rm h}^3} \{ \frac{\sigma_{\rm h}}{2} - \sqrt{\frac{\pi}{2}}|
\frac{\Im{\omega}}{k}|e^{\frac{(\Im{\omega})^2}{2 k^2 \sigma_{\rm h}^2}}
{\rm erfc} \left( \frac{\Im{\omega}}{\sqrt{2} k \sigma_{\rm h}} \right) \} \,,
\end{equation}
where erfc denotes the complementary error function. As is well known
perturbations will grow on length scales larger than the Jeans length
$\sqrt{\pi\sigma_{\rm h}^2/G \rho_{\rm b}} $.  However, dark halos are thought 
to be dynamically hot systems and their Jeans lengths will be of the order of
the size of the halos themselves. Thus this part of the solution of the 
Boltzmann equation (3) is uninteresting in the present context and will be 
not considered in the following.

\subsection{Response of the halo to a density wave in the disk}

I concentrate now on the remaining part of the Boltzmann equation (3),
\begin{eqnarray}
&&
\frac{\partial f_{\rm h1}}{\partial t} + u \frac{\partial f_{\rm h1}}
{\partial x} + v \frac{\partial f_{\rm h1}}{\partial y}
+ w \frac{\partial f_{\rm h1}}{\partial z} \\ &&
-\frac{\partial \Phi_{\rm d1}}{\partial x}\frac{\partial f_{\rm h0}}
{\partial u} \nonumber  - \frac{\partial \Phi_{\rm d1}}{\partial y}
\frac{\partial f_{\rm h0}}{\partial v} 
 - \frac{\partial \Phi_{\rm d1}}{\partial z}
\frac{\partial f_{\rm h0}}{\partial w} = 0 \nonumber \,,
\end{eqnarray}
which describes the halo response to a perturbation in the disk. If the
gravitational potential perturbation of the disk is Fourier expanded
the Fourier terms have the form (cf. equation 33 of Fuchs 2001)
\begin{equation}
\Phi_{\rm d{\bf k}_{||}}e^{i(k_{\rm x} x + k_{\rm y} y) - k_{\rm ||}|z|}
\end{equation}
with $k_{||}=|{\bf k}_{||}| =\sqrt{k_{\rm x}^2 + k_{\rm y}^2} $. This can be
converted to the form as in equation (9) by introducing Fourier coefficients
\begin{equation}
\Phi_{\rm d{\bf k}} = \frac{1}{2 \pi} \int_{-\infty}^{\infty} dz
\Phi_{\rm d{\bf k}_{||}} e^{-i k_{\rm z} z - k_{\rm ||}|z|} = \frac{1}{\pi}
\frac{k_{||}}{k_{||}^2 + k_{\rm z}^2}\Phi_{\rm d{\bf k}_{||}} \,.
\end{equation}
Note that the coordinate $y$, which is  defined in the reference
system of the disk, is related to the $y$ coordinate in the reference system of
the halo due to the motion of the center of the shearing sheet as
\begin{equation}
y \rightarrow y -r_{\rm 0} \Omega_{\rm 0} t \,.
\end{equation}
Fourier transforming the distribution function of the halo particles as $f_{\rm
h1} = \int d^3 k f_{\rm h {\bf k}} \exp{i ({\bf k},{\bf x})}$ and
changing again to $\xi, \eta, \zeta$ coordinates and the corresponding velocity
components the Boltzmann equation (16) takes the form
\begin{equation}
\frac{\partial f_{\rm h{\bf k}}}{\partial t} + \upsilon i k f_{\rm h{\bf k}} +
 i k \Phi_{\rm d{\bf k}} \frac{\upsilon}{\sigma_{\rm h}^2} f_{\rm h0} 
 e^{i(\omega - k_{\rm y}r_{\rm 0} \Omega_{\rm 0})t} = 0 \,,
\end{equation}
where $k=\sqrt{k_{||}^2 + k_{\rm z}^2}$. Equation (20) has been 
integrated with respect to the velocity components perpendicular to the 
$\xi$-axis. Assuming again a Gaussian velocity distribution for the halo 
particles it can be solved straightforward as
\begin{equation}
f_{\rm h{\bf k}} = - k \Phi_{\rm d{\bf k}} \frac{\upsilon}{\sigma_{\rm h}^2}
\frac{e^{i(\omega - k_{\rm y}r_{\rm 0} \Omega_{\rm 0})t}}
{\omega - k_{\rm y}r_{\rm 0} \Omega_{\rm 0} + k \upsilon}   \frac{\rho_{\rm b}}
{\sqrt{2 \pi}\sigma_{\rm h}}\exp{-\frac{\upsilon^2}{2 \sigma_{\rm h}^2}}\,.
\end{equation}
The corresponding Fourier term of the density perturbation of the halo
particles is given by
\begin{equation}
\rho_{\rm h{\bf k}} = - \Phi_{\rm d{\bf k}}\frac{\rho_{\rm b}}
{\sqrt{2 \pi}\sigma_{\rm h}^3} \int_{-\infty}^\infty d \upsilon
\frac{\upsilon e^{-\frac{\upsilon^2}{2 \sigma_{\rm h}^2}}}
{\frac{\omega}{k} -\frac{k_{\rm y}r_{\rm 0} \Omega_{\rm 0}}{k}
+ \upsilon} \,,
\end{equation}
where the time--dependent term has been split off and is written no longer
explicitely. Next the gravitational potential associated with this density 
distribution is calculated from the Poisson equation,
\begin{equation}
- k^2 \Phi_{\rm h{\bf k}} = 4 \pi G \rho_{\rm h{\bf k}}\,,
\end{equation}
resulting in
\begin{equation}
\Phi_{\rm h{\bf k}} = \Phi_{\rm d{\bf k}}\frac{4 \pi G \rho_{\rm b}}
{\sqrt{2 \pi}\sigma_{\rm h}^3} \frac{1}{k^2} \int_{-\infty}^\infty d \upsilon
\frac{\upsilon e^{-\frac{\upsilon^2}{2 \sigma_{\rm h}^2}}}
{\frac{\omega}{k} -\frac{k_{\rm y}r_{\rm 0} \Omega_{\rm 0}}{k}
+ \upsilon} \,.
\end{equation}
The integral on the rhs of equation (24) is not trivial and has to be evaluated
as a Cauchy principal value and the pole contribution at $\upsilon =
-\frac{\omega}{k} + \frac{k_{\rm y}r_{\rm 0} \Omega_{\rm 0}}{k}$, where
the halo particles travel at velocities in resonance with the Doppler-shifted
phase velocity of the wave. Following the method of Kegel (1998) I obtain for
the principal value of the integral
\begin{eqnarray}
&& \Phi_{\rm h{\bf k}}^{\rm nr}=\Phi_{\rm d{\bf k}}\frac{4 \pi G \rho_{\rm b}}
{\sigma_{\rm h}^2}\frac{1}{k^2} \{ 1 + i \sqrt{\pi} 
\frac{k_{\rm y}r_{\rm 0} \Omega_{\rm 0}-\omega}{\sqrt{2} k \sigma_{\rm h}} \\
&& \times {\rm erf} \left(i\frac{k_{\rm y}r_{\rm 0} \Omega_{\rm 0}-\omega}
{\sqrt{2} k \sigma_{\rm h}}\right) \exp{-\frac{(k_{\rm y}r_{\rm 0} 
\Omega_{\rm 0}- \omega)^2}{2 k^2 \sigma_{\rm h}^2}}\} \,, \nonumber 
\end{eqnarray}
where the error function with imaginary argument is related
to Dawson's integral. Since the gravitational forces in equation (1) have to be
taken at the midplane $z = 0$, it is necessary to convert
$\Phi_{\rm h{\bf k}}$ from the representation in {\bf k} space to a
mixed representation in $({\bf k}_{||}, z)$ space leading to
\begin{eqnarray}
&& \Phi_{\rm h{\bf k}_{||}}^{\rm nr}(z=0) = 
\int_{-\infty}^{\infty} d k_{\rm z}
\frac{k_{||}}{(k_{||}^2+ k_{\rm z}^2)^2} \Phi_{\rm d{\bf k}}\frac{4 G 
\rho_{\rm b}}{\sigma_{\rm h}^2} \nonumber \\ && \times \{ 1 + i \sqrt{\pi} \, 
\frac{k_{\rm y}r_{\rm 0} \Omega_{\rm 0}-\omega}{\sqrt{2} k \sigma_{\rm h}} \, 
{\rm erf} \left(i\frac{k_{\rm y}r_{\rm 0} \Omega_{\rm 0}-\omega}
{\sqrt{2} k \sigma_{\rm h}}\right) \\ && \times
\exp{-\frac{(k_{\rm y}r_{\rm 0} 
\Omega_{\rm 0}- \omega)^2}{2 k^2 \sigma_{\rm h}^2}}\} \nonumber \,.
\end{eqnarray}
The pole contribution to the integral in equation (24) has to be calculated
according to Landau's rule resulting in
\begin{eqnarray}
&& \Phi_{\rm h{\bf k}}^{\rm res}=\Phi_{\rm d{\bf k}}\frac{4 \pi G \rho_{\rm b}}
{\sigma_{\rm h}^2}\frac{1}{k^2} i \sqrt{\pi} \,
\frac{k_{\rm y}r_{\rm 0} \Omega_{\rm 0}-\omega}{\sqrt{2} k \sigma_{\rm h}} \\
&& \times  \exp{-\frac{(k_{\rm y}r_{\rm 0} 
\Omega_{\rm 0}- \omega)^2}{2 k^2 \sigma_{\rm h}^2}} \nonumber
\end{eqnarray}
and in the mixed representation
\begin{eqnarray}
&& \Phi_{\rm h{\bf k}_{||}}^{\rm res}(z=0) = - \int_{-\infty}^{\infty} 
d k_{\rm z} \frac{k_{||}}{(k_{||}^2+ k_{\rm z}^2)^2} \Phi_{\rm d{\bf k}}
\frac{i}{\sqrt{\pi}}\frac{4 \pi G \rho_{\rm b}}
{\sigma_{\rm h}^2} \nonumber \\ && \times  
\frac{\omega - k_{\rm y}r_{\rm 0} \Omega_{\rm 0}}{\sqrt{2} k \sigma_{\rm h}} \, 
\exp{-\frac{(k_{\rm y}r_{\rm 0} 
\Omega_{\rm 0}- \omega)^2}{2 k^2 \sigma_{\rm h}^2}} \,.
\end{eqnarray}
The final result can be formally written as
\begin{equation}
\Phi_{\rm h{\bf k}_{||}}(z=0) =\Upsilon (\omega - k_{\rm y}r_{\rm 0} 
\Omega_{\rm 0}, k_{||}) \Phi_{\rm d{\bf k}_{||}} \,,
\end{equation}
where the real and imaginary parts of $\Upsilon$ are defined by equations (26)
and (28), respectively. Thus for any given frequency there is a contribution
both from the non--resonant and the resonant halo particles. I find that in
dimensionless form
\begin{eqnarray}
&& \frac{(\omega - k_{\rm y}r_{\rm 0} 
\Omega_{\rm 0})^2}{2 (k_{||}^2 + k_{\rm z}^2)\sigma_{\rm h}^2} = 
\\ && 1.748 \left(
\frac{\sigma_{\rm d}}{\sigma_{\rm h}} \right)^2 \frac{1}{Q^2}
\frac{k_{\rm crit}^2}{k_{||}^2+ k_{\rm z}^2}
\frac{(\omega - k_{\rm y}r_{\rm 0} \Omega_{\rm 0})^2}{\kappa^2} \,, \nonumber
\end{eqnarray}
where $\sigma_{\rm d}$ denotes the velocity dispersion of the disk stars and 
$k_{\rm crit}= 2\pi/\lambda_{\rm crit}$, the critical wave number 
of the disk. The function $\Upsilon$ is illustrated in Fig.~1 for parameters 
typical for the solar neighbourhood in the Milky Way. The velocity dispersion of
the halo particles has been estimated as 
$\sigma_{\rm h} = r_0\Omega_0 /\sqrt{2}$ like in an isothermal sphere.
\begin{figure} [h]
\begin{center}
\epsfxsize=8.5cm
   \leavevmode
     \epsffile{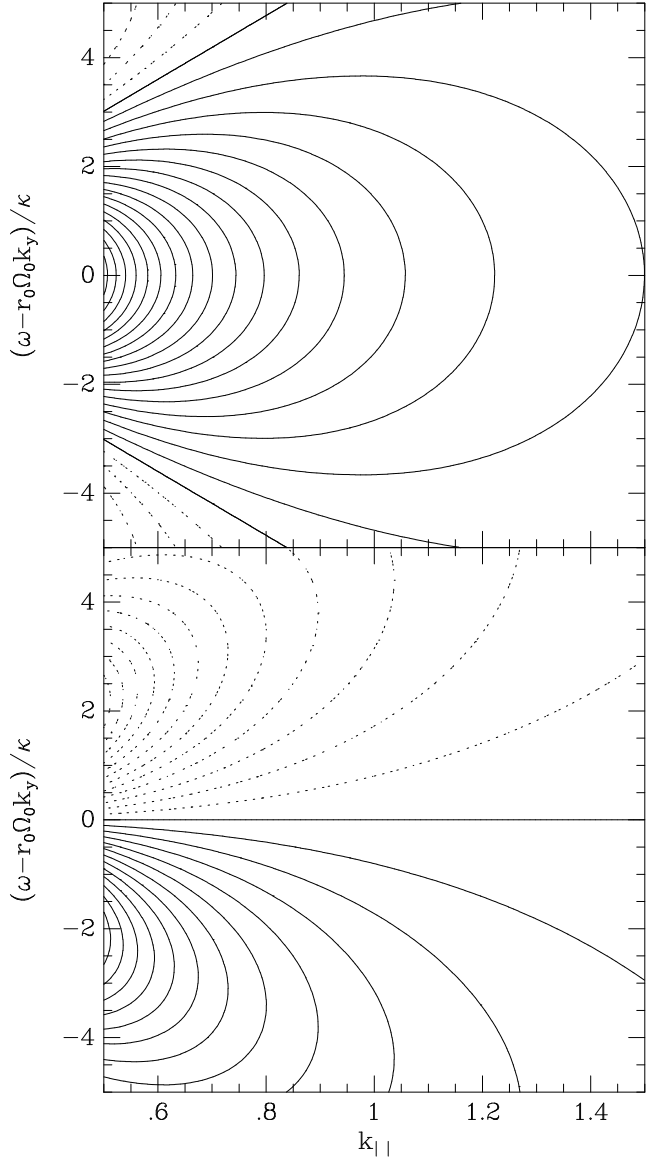}
\caption{The function $\Upsilon (\omega - k_{\rm y}r_{\rm 0} 
\Omega_{\rm 0}, k_{||})$. The real part is shown in the upper panel, while 
the imaginary part is shown in
the lower panel. The Doppler shifted frequency is given in terms of the
epicyclic frequency and the wave number $k_{||}$ in units of the critical wave
number. Contours are given at levels of $ 10^{-2} G \rho_{\rm b}/\kappa^2$.
Negative values are indicated as dotted lines. A $Q$ parameter value of 
$Q = 1.4$ and a ratio of disk to halo velocity dispersions of 
$\sigma_{\rm d} : \sigma_{\rm h} = 1 : 5$ are assumed. }
\label{fig1}
   \end{center}
   \end{figure}

\section{Disk dynamics} 

The halo response (29) to the perturbation in the disk has to be inserted into
equation (1). Solving the Boltzmann equation (1) is greatly facilitated by the
fact that its form is identical to the case of an isolated shearing sheet with
the replacement
\begin{equation}
\Phi_{\rm d{\bf k}} \rightarrow (1+\Upsilon)\Phi_{\rm d{\bf k}}
\end{equation}
and I can use directly the results of Fuchs (2001) even though the Boltzmann
equation is treated there using action and angle variables instead of the
Cartesian coordinates as in equation (1). In particular the factor
$1 + \Upsilon$ is carried straightforward through to the fundamental Volterra
integral equation (equation 68 of Fuchs 2001)
\begin{eqnarray}
\Phi_{\rm {\bf k'},\omega}& = &\int_{-\infty}^{k'_{\rm{x}}} dk_{\rm{x}} 
\mathcal{K} 
\left(k_{\rm{x}},k'_{\rm{x}}\right) (1+\Upsilon(k_{\rm x},k'_{\rm y},\omega))
\Phi_{\rm k_{\rm x}, k'_{\rm y}, \omega} 
\\ & + & r_{\rm{\bf{k'},\omega}} \nonumber\,,
\end{eqnarray}
where the kernel $\mathcal{K}$ is given by equation (67) of Fuchs (2001).  
$r_{\rm{\bf{k'}}}$ describes an inhomogeneity of equation (32) related  
to an initial non--equilibrium state of the shearing sheet. Equation (32) is
separating in the circumferential wave number $k'_{\rm y}$. 
In equation (32) the wave numbers are expressed in units of the critical wave
number $k_{\rm crit}$ (cf.~equation 30). This implies that the Volterra equation
describing a shearing sheet embedded in a rigid halo potential is formally the
same as that of an isolated shearing sheet, because in this case $\Upsilon = 0$
and the halo mass affects only the numerical values of the critical wave number 
$k_{\rm crit}$ and the stability parameter $Q$. It is
advantageous to consider equation (32) transformed back from frequency to time
domain. Splitting off the $\omega$--dependent term $\exp{i \omega  
\frac{k_{\rm x}-k'_{\rm x}}{2 A k'_{\rm y}}}$ from the kernel and making use 
of the convolution theorem of the Fourier transform of products of two
functions leads to
\begin{eqnarray}
 \Phi_{\rm {\bf k'},t} & = &\int_{-\infty}^{k'_{\rm{x}}} dk_{\rm{x}} 
\tilde{\mathcal{K}}\left(k_{\rm{x}},k'_{\rm{x}}\right) \nonumber \\ &\times& \{
\int_{0}^{\infty} dt' \Phi_{\rm k_{\rm x}, k'_{\rm y},t'} \delta
\left( t-t'+ \frac{k_{\rm x}-k'_{\rm x}}{2 A k'_{\rm y}} \right) \\ &+&
\int_{0}^{\infty} dt' \Phi_{\rm k_{\rm x}, k'_{\rm y}, t'} \mathcal{F}
\left(\Upsilon(k_{\rm x},k'_{\rm y},\omega)
e^{i\omega\frac{k_{\rm x}-k'_{\rm x}}{2 A k'_{\rm y}}}\right)_{\rm t-t'} \}
\nonumber \\ & + & r_{\rm {\bf k'},t}\,,\nonumber
\end{eqnarray}
where the operator $\mathcal{F}$ denotes the Fourier transform from $\omega$ 
to time domain. In equation (33) I have assumed an initial perturbation of the 
disk at time $t = 0$ so that $\Phi_{{\rm k_{\rm x}, k'_{\rm y}},t'<0}=0$. The 
Fourier transform $\mathcal{F}$ can be calculated
analytically using formulae 6.317\footnote{I use the identity $\int_0^\infty d
\omega \cos{(b\omega)} a \omega {\rm erf}(i a \omega)e^{-a^2\omega^2} = a
\frac{\partial}{\partial b}\int_0^\infty d\omega \sin{(b\omega)}{\rm erf}
(i a \omega)e^{-a^2\omega^2}$ with formula 6.317.} and 3.952, and the 
integrals with respect to $k_{\rm z}$ using formula 3.466 of Gradshteyn 
\& Ryzhik (2000) leading to
\begin{eqnarray}
&& \mathcal{F}\left(\Upsilon(k_{\rm x},k'_{\rm y},\omega)
e^{i\omega\frac{k_{\rm x}-k'_{\rm x}}{2 A k'_{\rm y}}}\right)_{\rm t-t'} = 
\nonumber \\ && \frac{4 \pi^2 G \rho_{\rm b}}{\sigma_{\rm h}^2}\frac{1}{k_{\rm
x}^2+{k'_{\rm y}}^2} 
\delta \left( t-t'+ \frac{k_{\rm x}-k'_{\rm x}}{2 A k'_{\rm y}} \right) \\ &&
+ 4 \pi^2G \rho_{\rm b}\exp{[ik'_{\rm y}r_{\rm 0}\Omega_{\rm 0}\left( t-t'+ 
\frac{k_{\rm x}-k'_{\rm x}}{2 A k'_{\rm y}} \right)]} \nonumber \\ && \times
\{\left( t-t'+ \frac{k_{\rm x}-k'_{\rm x}}{2 A k'_{\rm y}} \right) + \Big|
t-t'+ \frac{k_{\rm x}-k'_{\rm x}}{2 A k'_{\rm y}} \Big| \} \nonumber \\ &&
\times {\rm erfc} \left( \frac{\sigma_{\rm h}}{\sqrt{2}} \sqrt{k_{\rm x}^2 
+{k'_{\rm y}}^2}\Big|t-t'+ \frac{k_{\rm x}-k'_{\rm x}}{2 A k'_{\rm y}} 
\Big| \right)\nonumber \,,
\end{eqnarray}
where the terms depending on the delta function and the absolute value
$\Big|t-t'+ \frac{k_{\rm x}-k'_{\rm x}}{2 A k'_{\rm y}} \Big|$ in the curly 
bracket represent the contribution from the non-resonant halo particles, while 
the remaining term gives the contribution by the resonant halo particles. If 
this is inserted into equation (33) it takes the form
\begin{eqnarray}
 && \Phi_{\rm {\bf k'},t} = \int_{-\infty}^{k'_{\rm{x}}} dk_{\rm{x}} 
\tilde{\mathcal{K}}\left(k_{\rm{x}},k'_{\rm{x}}\right) \{
\Phi_{\rm k_{\rm x}, k'_{\rm y},t+\frac{k_{\rm x}-k'_{\rm x}}{2 A k'_{\rm y}}}
 \\ && +
\int_{0}^{t+\frac{k_{\rm x}-k'_{\rm x}}{2 A k'_{\rm y}}} dt'
 \Phi_{\rm k_{\rm x}, k'_{\rm y}, t'} \mathcal{F}
\left(\Upsilon(k_{\rm x},k'_{\rm y},\omega)
e^{i\omega\frac{k_{\rm x}-k'_{\rm x}}{2 A k'_{\rm y}}}\right)_{\rm t-t'} \}
\nonumber \\ && + r_{\rm {\bf k'},t} \,. \nonumber 
\end{eqnarray}
Equation (35) can be integrated numerically with very modest numerical effort.
In Fig.~2 I illustrate the response of the shearing sheet now embedded in a live
halo to an initial sinusoidal perturbation of unit amplitude. For this purpose 
I use the inhomogeneity term of the Volterra equation
\begin{equation}
r_{\rm {\bf k'},\omega} = \int_{-\infty}^{k'_{\rm{x}}} dk_{\rm{x}} 
{\mathcal{L}}\left(k_{\rm{x}},k'_{\rm{x}}\right)
f_{\rm k_{\rm x}, {k'}^{\rm in}_{\rm y}}(0)
\end{equation}
derived in Fuchs (2001) with $f_{\rm k_{\rm x}, {k'}^{\rm in}_{\rm y}}(0) 
\propto \delta ( k_{\rm x} -k_{\rm x}^{\rm in}) $. The response of the 
shearing sheet to this initial impulse is a swing amplification event. The
radial wave number $k_{\rm x}$ evolves as
\begin{equation}
k_{\rm x}=k_{\rm x}^{\rm in} +2 A {k'_{\rm y}}^{\rm in} t \,,
\end{equation}
while the circumferential wave number ${k'_{\rm y}}^{\rm in} $ is constant,
which means that the wave crests swing around from leading to trailing 
orientation during the amplification phase. Around $t = 6$ the amplitudes 
become negative which indicates that the swing amplified density wave is also 
oscillating. As can be seen from Fig.~2 comparing the evolution of 
shearing sheets embedded either in a rigid halo potential or a live halo this
characteristic behaviour of the density wave is not changed by the responsive 
halo, but the maximum growth factor of the amplitude of the wave is enhanced by
a surprisingly large amount. In Fig.~2 the evolution of a
shearing sheet embedded in a static halo potential has been calculated from
equation (35) setting $\mathcal{F} = 0$. The maximum growth factor is identical
to within 1\% accuracy with the maximum growth factor calculated by Fuchs (2001)
in an alternative way. In order to test the accuracy of the numerical solution 
of the full equation (35) I have increased the velocity dispersion of the halo
particles to $\sigma_{\rm d} : \sigma_{\rm h} = 1 : 13$. The halo is then very
stiff and the enhancement of the maximum growth factor of the amplitude of
the wave by the responsive halo is only 5.3\% so that the solution of equation
(35) is very similar to that for a density wave in a shearing sheet in a static
halo. If I use this, which is correctly calculated, in
the second term of equation (35) I confirm by this iterated solution the maximum
growth factor of the self--consistent solution of equation (35) to an accuracy
better than 1\%.
\begin{figure} [h]
\begin{center}
\epsfxsize=8.5cm
   \leavevmode
     \epsffile{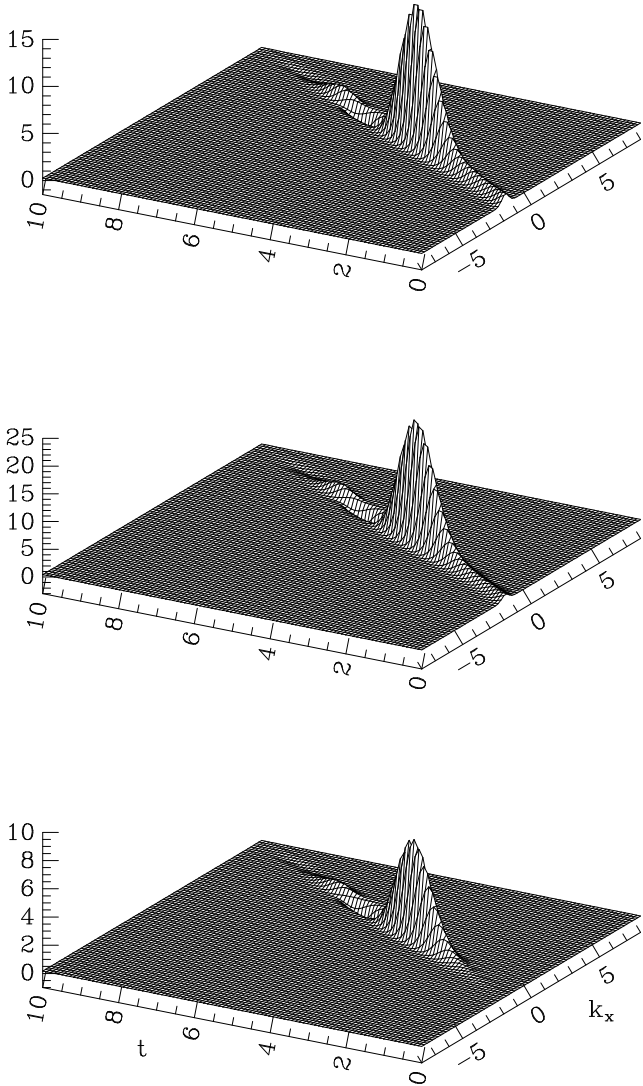}
\caption{Swing amplified density wave in the shearing sheet. The upper diagram
shows the evolution in a shearing sheet embedded 
in a static halo potential triggered by an impulse with 
unit amplitude and wave vector ${\bf k}^{\rm in}=(-2,0.5)k_{\rm crit}$.  
Time is given in units of 0.045 epicycle periods. The middle diagram shows the
evolution of a shearing sheet embedded in a live dark halo triggered by the
same impulse. The lower diagram  shows the difference. The model parameters are
$A/\Omega_0=0.5, Q=1.4, \sigma_{\rm d}: \sigma_{\rm h}=1:5, G\rho_{\rm
b}/\kappa^2=0.01$, and $r_0\Omega_0:\sigma_{\rm d}=220:44$.}
\label{fig2}
   \end{center}
   \end{figure}
The enhanced maximum growth factor of swing amplified density
waves due to a responsive halo seems to be
the equivalent of the enhanced growth of bars of stellar disks embedded in live
dark halos seen in the numerical simulations. However, the interaction of the 
shearing sheet and the surrounding halo is not only mediated by the resonant 
halo particles, but the non--resonant halo particles play an important role as
well. This can be demonstrated, for instance, by keeping in equation (34)
only that part of the $\Upsilon$ function related to the resonant halo 
particles and solving equation (35) approximately by iterating the solution for 
the isolated sheet,
\begin{eqnarray}
\Phi_{\rm {\bf k'},t} &\approx & \int_{-\infty}^{k'_{\rm{x}}} dk_{\rm{x}} 
\tilde{\mathcal{K}}\left(k_{\rm{x}},k'_{\rm{x}}\right) \{
\Phi_{\rm k_{\rm x}, k'_{\rm y},t+\frac{k_{\rm x}-k'_{\rm x}}{2 A k'_{\rm y}}}
\\ & + &
\int_{0}^{\infty} dt' \Phi_{\rm k_{\rm x}, k'_{\rm y}, t'}^{\rm iso} \mathcal{F}
\left(\Upsilon^{\rm res}(k_{\rm x},k'_{\rm y},\omega)
e^{i\omega\frac{k_{\rm x}-k'_{\rm x}}{2 A k'_{\rm y}}}\right)_{\rm t-t'} \}
\nonumber \\ &+& r_{\rm {\bf k'},t} \nonumber \,.
\end{eqnarray}
   
The amplification of density waves depends critically on the Toomre stability
parameter $Q$. This is illustrated in Fig.~3 where the response of the shearing
sheet to the same initial impulse as in the previous example is shown, but
assuming a stability parameter of $Q=2$. As can be seen in Fig.~3 there is
neither effective amplification of density waves in a shearing sheet in a rigid
halo potential or in a shearing sheet embedded in a live dark halo.
\begin{figure} [h]
\begin{center}
\epsfxsize=8.5cm
   \leavevmode
     \epsffile{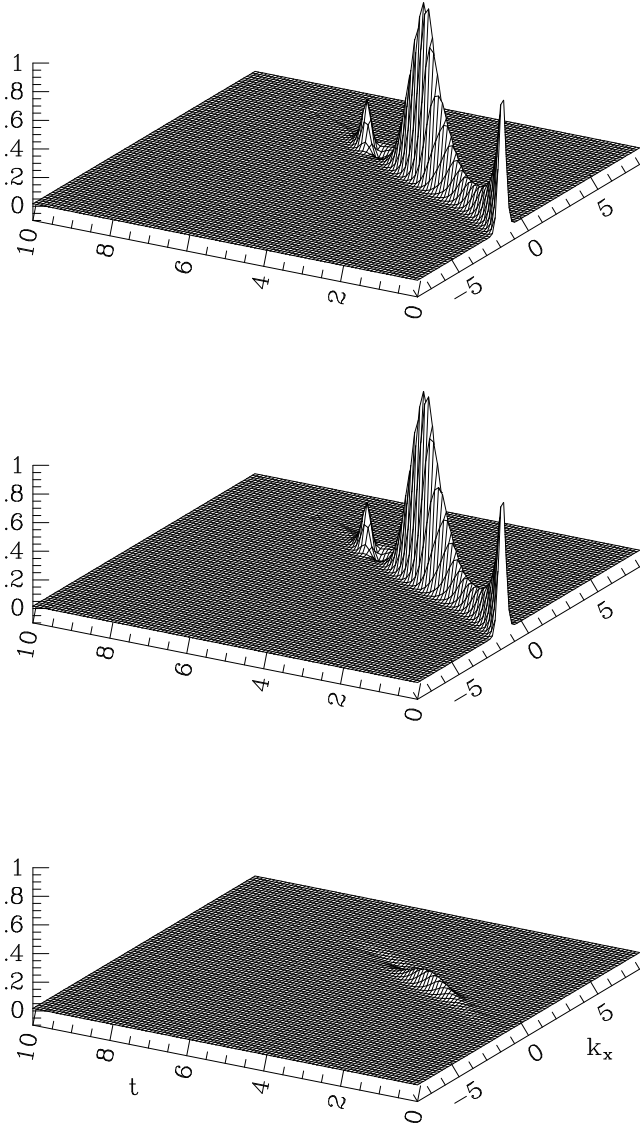}
\caption{Same as in Fig.~2, but adopting $Q=2$.}
\label{fig3}
   \end{center}
   \end{figure}

In Fig.~4 I illustrate the maximum growth factors as function of the
circumferential wave number $k_{\rm y}$ for shearing sheets embedded in a rigid
halo potential or a live dark halo, respectively. As can be seen from Fig.~4 in
both cases maximum amplification occurs for wave numbers around 
$k_{\rm y} = 0.5$ or in the notation of Toomre (1981)
$X = k_{\rm y}^{-1} = 2$. Thus the preferred circumferential wave length of 
the density waves is $2 \lambda_{\rm crit}$.
\begin{figure} [h]
\begin{center}
\epsfxsize=8.5cm
   \leavevmode
     \epsffile{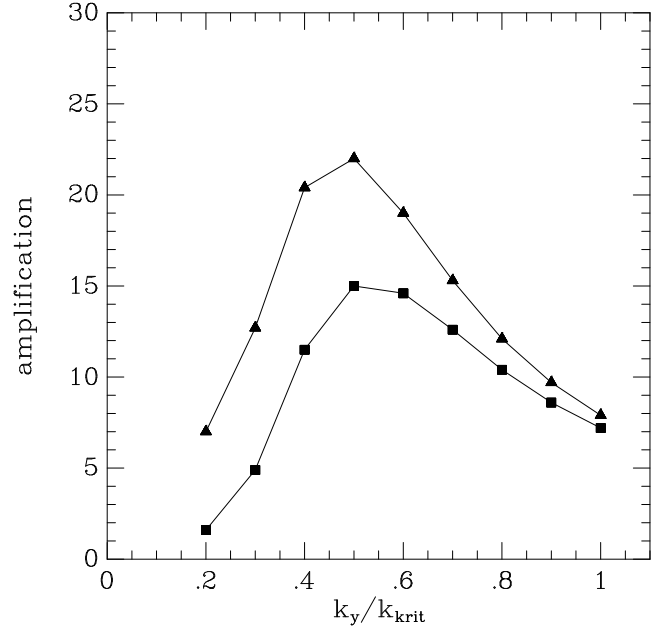}
\caption{Maximum growth factors of swing amplified density waves in a 
shearing sheet embedded in a rigid halo potential (squares) and in a shearing 
sheet embedded in a live dark halo (triangles). The same model parameters as 
above have been adopted.}
\label{fig4}
   \end{center}
   \end{figure}
   
Following Athanassoula et al.~(1987) arguments of swing amplification theory 
have been used to constrain the mass--to-light ratios of galactic disks. The 
expected number of spiral arms is determined by how often the azimuthal wave 
length $2\pi/k_{\rm y}$ fits onto the circular annulus,
\begin{equation}
m = \frac{2 \pi r_0}{\frac{2 \pi}{k_{\rm y}}}\,.
\end{equation}
As shown in Fig.~4 density waves grow preferentially with an azimuthal wave
number around $k_{\rm y} \approx 0.5 k_{\rm crit}$. This assumes a disk with a
flat rotation curve, $A/\Omega_0=0.5$, but the preferred wave length
varies with the slope of the rotation
curve measured by Oort's constant $A$ (cf.~Fuchs 2001). The observed number of
spiral arms of a galaxy taken together with the rotation curve allows then to
determine the surface density of the disk. Since the wave number of maximum
growth of the density wave amplitudes is the same for shearing sheets embedded 
either in a rigid halo potential or in a live dark halo, the presence of a 
responsive dark halo does not influence the constraints on the disk mass 
obtained this way.

Another constraint on the disk mass depends on the implied value of the Toomre 
stability parameter $Q$ (Fuchs 1999, 2003). $Q$ must be larger than $Q=1$ in 
order to avoid Jeans like instabilities in the disks (Toomre 1964). A disk 
violating this condition will evolve rapidly by fierce dynamical instabilities 
above the threshold of $Q=1$ (Fuchs \& von Linden 1998). On the other hand, I 
have demonstrated above that the $Q$ value must be less than $Q=2$ in order 
that the disk can develop appreciable spiral structure. If the radial velocity
dispersions of the stars are known, the allowed range of $Q$ sets constraints on 
the disk mass, which are independent from the previous ones (cf.~equations 6
and 7). As shown in Fig.~3 this condition is not changed by the presence of a 
live dark halo.

\section{Momentum transfer to the halo}

The analogue of the angular momentum transfer from bars to dark halos is in the
shearing sheet the transfer of linear momentum to the surrounding halo
particles. In order to describe this I adapt the discussion of Dekker (1976) to
the shearing sheet model. The acceleration of a halo particle is given by
\begin{equation}
\dot{{\bf v}} = -\nabla \Phi \,,
\end{equation}
where I assume a gravitational potential of the form
\begin{equation}
\Phi({\bf x}, t)= \Re{ \{ \Phi({\bf x}) e^{i \omega t} \} }\,,
\end{equation}
with complex frequency $\omega$. Thus the overall acceleration of the halo is
given by
\begin{equation}
<\dot{{\bf v}}> = -\int d^3x \int d^3v f_{\rm h} \nabla \Phi  \,.
\end{equation}
If both the gravitational potential and the distribution function of the halo
particles are Fourier expanded\footnote{It can be shown (Binney \& Tremaine
1987 and references therein) that if the Fourier transformed
potential of a point mass is inserted into equation (43) this leads
exactly to Chandrasekhar's dynamical friction formula, although with a
Coulomb logarithm defined in a slightly different way.},
\begin{eqnarray}
<\dot{{\bf v}}> &=& -\int d^3x \int d^3v \int d^3k\,i {\bf k}
\Phi_{\rm {\bf k}}(t) e^{i({\bf k},{\bf x})} \nonumber \\ 
&\times& \int d^3k' f_{\rm h{\bf k'}}(t) 
e^{i({\bf k'},{\bf x})} \\ &=& (2 \pi)^3 \int d^3v \int d^3k\, i {\bf k}
\Phi_{\rm {\bf -k}}(t) f_{\rm h{\bf k}}(t) \nonumber \,.
\end{eqnarray}
In order that the gravitational potential $\Phi$ be a real quantity its Fourier
coefficients must obey the relation $\Phi_{\rm {\bf -k}}(t)=\Phi_{\rm {\bf k}}^*
(t)$, and for a potential of the form (41) one can write
\begin{equation}
\Phi_{\rm {\bf k}}(t)=\frac{1}{2} \{\Phi_{\rm {\bf k}}e^{i \omega t}+ 
\Phi_{\rm {\bf -k}}^*e^{-i \omega^* t}\}\,.
\end{equation}
Again spatial coordinates $\xi, \eta,$ and $\zeta$ are introduced with the
$\xi$--axis parallel to the wave vector ${\bf k}$ and using equation (11)
one obtains from equation (43)
\begin{eqnarray}
<\dot{\upsilon}>&=& i (2 \pi)^3 \int d^3k \int_{-\infty}^{\infty} d \upsilon
\frac{\rho_{\rm b}}{\sqrt{2 \pi} \sigma_{\rm h}^3} e^{-\frac{\upsilon^2}
{2\sigma_{\rm h}^2}} \nonumber \\ &\times& \frac{k}{4}\{\Phi_{\rm {\bf -k}}
e^{i \omega t}+\Phi_{\rm {\bf k}}^*e^{-i \omega^* t}\} \\ &\times&
\{-k \Phi_{\rm {\bf k}}\frac{\upsilon e^{i \omega t}}{\omega+k \upsilon}
-k\Phi_{\rm {\bf -k}}^*\frac{\upsilon e^{-i \omega^* t}}{-\omega^* + k 
\upsilon} \} \nonumber \,.
\end{eqnarray}
In equation (45) I have integrated already over the velocity components
perpendicular to the $\xi$-axis. Two of the four terms of the integrand 
cancel out leading to
\begin{eqnarray}
<\dot{\upsilon}>&=& - 2 \pi^3 i \int d^3k \int_{-\infty}^{\infty} d \upsilon
\frac{\rho_{\rm b}}{\sqrt{2 \pi} \sigma_{\rm h}^3} e^{-\frac{\upsilon^2}
{2\sigma_{\rm h}^2}} \\ &\times& k^2 e^{-2 \Im{\omega} t}
|\Phi_{\rm {\bf k}}|^2
\{\frac{\upsilon }{\omega+k \upsilon}
-\frac{\upsilon }{\omega^* + k\upsilon} \} \nonumber \,.
\end{eqnarray}
In the limit of $-\Im{\omega}\rightarrow 0$
\begin{equation}
\frac{\upsilon }{\omega+k \upsilon}
-\frac{\upsilon }{\omega^* + k\upsilon}\rightarrow - i 2 \pi \upsilon
\delta(\Re{\omega}+k \upsilon) \,,
\end{equation}
and I find
\begin{eqnarray}
<\dot{\upsilon}>&=& -4 \pi^4 \int d^3k \frac{\rho_{\rm b}}
{\sqrt{2 \pi} \sigma_{\rm h}^3} |\Phi_{\rm{\bf k}}|^2
 (\omega-k_{\rm y}r_0\Omega_0) \\ &\times&  
 \exp{-\frac{(\omega-k_{\rm y}r_0\Omega_0)^2}{2 k^2 \sigma_{\rm h}^2}} \,,
 \nonumber 
\end{eqnarray}
where $\omega$ denotes now the frequency of the wave in the reference frame of
the shearing sheet as in the previous sections. Note that according to the
notation used here $(\omega-k_{\rm y}r_0\Omega_0) < 0$ for waves traveling 
along the $y$-axis in the forward direction, so that
$<\dot{\upsilon}>\,\, > 0$ for 
such waves. From the form of the Fourier coefficients (18) it becomes 
immediately clear that only planar momentum is transferred to the halo, 
$< \dot{v}_{\rm z} > = 0$, as expected from symmetry reasons.
 $< \dot{v}_{\rm x} > $ and $< \dot{v}_{\rm y} > $ depend on the exact form of
 the power spectrum $|\Phi_{\rm{\bf k}}|^2 $. Equation (47) shows that 
linear momentum transfer to the halo is mediated entirely by halo particles on 
orbits resonant with the wave. This is exactly the same as for angular momentum
transfer from bars to halos (Weinberg 1985, Weinberg \& Katz 2002, Athanassoula
2003, Sellwood 2003). It has also been known for a long time for the angular 
momentum exchange between spiral density waves and galactic disks 
(Lynden-Bell \& Kalnajs 1972) or for momentum transfer from longitudinal plasma
waves (Stix 1962). It is interesting that the exchanged linear momentum scales 
as $\rho_{\rm b}/ \sigma_{\rm h}^3$ in equation (48). This implies that at given
mass a dynamical hot halo absorbs less momentum than a cooler one. On the other 
hand, the shearing sheet would loose less momentum. The same effect has been 
described by Athanassoula (2003) for the angular momentum transfer from bars to 
halos. 

\section{Summary and conclusions}

If a self--gravitating shearing sheet is embedded in a live dark halo, the halo
particles respond unexpectedly strong to density waves in the sheet. The
interaction between the density waves and the halo particles is mediated 
both by halo particles on orbits in resonance with the waves and on
non--resonant orbits. If the embedded shearing sheet is initially perturbed by
a sinusoidal wave, a swing amplified density wave develops in the disk, which 
is of the same type as in an isolated sheet or a sheet in a static
halo potential, but with an amplitude
enhanced by a surprisingly large amount. This appears to be the equivalent of 
the enhanced bar growth in stellar disks embedded in live dark halos instead of
static halo potentials seen in numerical simulations. There is transfer of
linear momentum of density waves in the shearing sheet to the halo particles.
This is mediated, however, entirely by halo particles on orbits in resonance 
with the waves similar to the torque exerted by bars on the surrounding halo.

\acknowledgements{I thank E. Athanassoula for stimulating discussions and the
anonymous referee for helpful comments.}

{}

\end{document}